\documentclass[aps,prd,preprint,superscriptaddress]{revtex4}
\usepackage{graphicx}
\usepackage{amsfonts,amsmath,amssymb,bm}

\begin{document}

\title{Shear viscosity from thermal fluctuations in relativistic conformal fluid dynamics}

\author{J. Peralta-Ramos}
\email{jperalta@ift.unesp.br}
\affiliation{Instituto de F\'isica Te\'orica, Universidade Estadual Paulista, 
Rua Doutor Bento Teobaldo Ferraz 271 - Bloco II, 01140-070 S\~ao Paulo, Brazil}
\affiliation{Departamento de F\'isica, FCEyN-UBA, and IFIBA-CONICET, Ciudad Universitaria, Pabell\'on I, Buenos Aires 1428, Argentina}

\author{E. Calzetta}
\email{calzetta@df.uba.ar}
\affiliation{Departamento de F\'isica, FCEyN-UBA, and IFIBA-CONICET, Ciudad Universitaria, Pabell\'on I, Buenos Aires 1428, Argentina}

\date{\today}

\begin{abstract}
Within the framework of relativistic fluctuating hydrodynamics we compute the contribution of thermal fluctuations to 
the effective infrared shear viscosity of a conformal fluid, focusing on quadratic (in fluctuations), second order (in velocity gradients) terms in the conservation equations. Our approach is based on 
the separation of hydrodynamic fields in soft and ultrasoft sectors, in which the effective shear viscosity arises due to the action of the soft modes on the evolution of the ultrasoft ones. We find that for a strongly coupled fluid with small shear viscosity--to--entropy ratio $\eta/s$ the contribution of thermal fluctuations to the effective shear viscosity is small but significant. Using realistic estimates for the strongly coupled quark--gluon plasma created in heavy ion collisions, we find that for $\eta/s$ close to the AdS/CFT lower bound $1/(4\pi)$ the correction is positive and at most amounts to $10 \%$ in the temperature range $200$--$300$ MeV, whereas for larger values $\eta/s \sim 2/(4\pi)$ the correction is negligible. For weakly coupled theories the correction is very small even for $\eta/s=0.08$ and can be neglected.
\end{abstract}

\maketitle

\section{Introduction}
The heavy ion collisions experiments performed at RHIC and LHC 
create a hot and dense medium, the so-called Quark-Gluon plasma (QGP), that 
is believed to be strongly coupled. The most compelling evidence supporting 
this idea comes from relativistic viscous fluid dynamics simulations 
that reproduce the momentum anisotropy patterns measured by experiment (see, e.g. Refs. \cite{exp1,exp2,exp3,exp4,rev1,rev2,rev3,ext1,ext2,ext3,ext4,PRC-E}). The momentum anisotropy 
is the translation to momentum space of the initial spatial eccentricity of non central collisions. 
A weakly interacting system of particles can not convert spatial anisotropy into momentum anisotropy in an efficient way, but this translation 
can occur quite efficiently if the particles interact strongly. The observed large magnitude of the momentum anisotropy indicates that the QGP is strongly coupled. Another indication of the strongly coupled nature of the QGP is the 
strong quenching of high-$p_T$ probes measured at RHIC \cite{exp1,exp2,exp3,exp4}. 

Transport coefficients such as the shear ($\eta$) and bulk ($\zeta$) viscosities are crucial inputs in fluid dynamics simulations attempting to describe the evolution of matter created in heavy ion collisions.   
Great efforts are currently focused on developing new theoretical tools to compute these transport coefficients more accurately from microscopic models  \cite{tr1,ttr1,ttr2,ttr3,ttr4,ttr5,ttr6,ttr7,ttr8,ttr9,ttr10,ttr11,hyd1,hyd2,hyd3,deni}, and also on extracting them more precisely from RHIC and LHC measurements \cite{rev1,rev2,rev3,ext1,ext2,ext3,ext4}.  Results obtained from Lattice QCD calculations \cite{tr1,ttr6,lat,lat2} indicate that $\eta/s$ and $\zeta/s$, where $s$ is the entropy density, depend significantly on temperature, showing a minimum and maximum, respectively, at the critical temperature corresponding to the QGP--hadron crossover. Note that although it is expected that the transport coefficients of the QGP depend on temperature, the impact of a temperature--dependent $\eta/s$ on momentum anisotropies as obtained from simulations has been investigated only recently \cite{temp1,temp2,temp3,temp4,PRC-G}. 

It is possible to extract an average value of $\eta/s$ by matching the particle spectra and elliptic flow obtained from fluid dynamics \cite{ext1} or hybrid fluid--kinetic simulations \cite{ext2,ext3,ext4} to data. See also Refs. \cite{ext5,ext6,ext7} for a description of other approaches to extract values of $\eta/s$ from data. The optimal value of $\eta/s$ that comes out of all these fits is close to the 
lower bound $\eta/s = 1/(4\pi)$ that was found by Kovtun, Son and Starinets \cite{KSS} (KSS bound from now on) via the Anti de Sitter/Conformal Field Theory (AdS/CFT) correspondence. However, deviations from the KSS bound by a factor of two or more are still possible due to uncertainties stemming from diverse sources such as the initial conditions, the dynamics of chiral fields coupled to the quark fluid, and 
the freeze out stage, to mention a few -- see Refs. \cite{rev1,rev2,rev3,ext1,ext2,ext3,ext4,PRC-G,PRC-E,unc1,unc2,unc3,PRD-link,dus,luz}. 
Very low values of $\eta/s$ are also favored by studies of the overall entropy production during a collision -- see e.g. \cite{entropy} -- 
and, albeit more indirectly, by the fact that flow structures such as cones and ridges are able to survive until freeze out. If dissipative effects were too large, such collective flow structures would be washed away much earlier in the evolution of the fireball \cite{struc1,struc2,struc3,struc4}. 

The system created in heavy ion collisions consists of a fireball of quarks and gluons that expands and cools very rapidly under its own pressure. Since the fireball is formed out of a finite number of particles and has a small size in the range $5$--$10$ fm, it is natural to expect that thermal fluctuations may have an observable impact on its evolution. In spite of this expectation, the role of thermal fluctuations and its impact on hydrodynamic evolution has only very recently been discussed in the context of heavy ion collisions \cite{kovtun,lub,kap}.

It was shown in Ref. \cite{kovtun} that for the QGP with small values of $\eta/s$ near the KSS bound, thermal fluctuations on top of long--lived sound and shear waves can contribute significantly to the effective value of $\eta/s$ as would be measured on large scales. In contrast, for larger values $\eta/s \sim 2/(4\pi)$ the correction coming from thermal fluctuations on sound and shear waves was found to be negligible.  
Specifically, in \cite{kovtun} the authors used Kubo's formula together with second order hydrodynamics and focused on corrections to $\eta/s$ coming from the zeroth order (in gradients), quadratic term $(\rho_0 + P_0)u^\mu u^\nu$ in the energy momentum tensor $T^{\mu\nu}$, where 
$\rho_0$ and $P_0$ are the equilibrium energy density and pressure ($P_0 = \rho_0/3$ for a conformal fluid), respectively, and $u^\mu$ is the flow velocity fluctuation (\cite{kovtun} considers a vanishing background velocity). 
A related study on the effective shear viscosity of the strongly coupled QGP was carried out in \cite{lub} using the AdS/CFT correspondence, where corrections to $\eta$ arising from higher order velocity gradients were computed using the idea of resumming those corrections into an effective $\eta(\omega,k)$ depending on both frequency and momentum. 

In this work we compute the correction to the shear viscosity that come from the effect of relatively small scale fluctuations of the hydrodynamic modes on their evolution on larger scales. For this purpose, we use the equations of relativistic fluctuating fluid dynamics \cite{landau,fox,esteban-fluc,librofluc,libro,kap} and focus on terms which are second order in velocity gradients (in $T^{\mu\nu}$ these terms are first order) and quadratic in velocity fluctuations. The development presented here is based on a division of hydrodynamic modes in ultrasoft and soft sectors, both of them subjected to the influence of stochastic noise coming from hard (i.e. particle--like) modes. It is, therefore, inspired in the hard thermal loops approach to thermal field theory \cite{LeB96,BraPis90a,FreTay90} and particularly in earlier studies of the Langevin dynamics of soft and ultrasoft modes \cite{Bod98,Bod99,ArSoYa99a,ArSoYa99b,LitMan02}. As this approach does not rely on Kubo's formula it is different from the one adopted in \cite{kovtun}. We anticipate that the correction computed here is of order $\chi^2 \equiv p^2_{max}\eta^2/(sT)^2$ or higher, where $p_{max}$ is the value of momentum beyond which the second order gradient expansion breaks down, while the correction obtained in \cite{kovtun} is of order $\chi$. Since $\chi$ must be small for a fluid description to be reliable, the correction to $\eta/s$ obtained here is suppressed with respect to the one of \cite{kovtun}, but as we will show it is still significant. We will compare our results with those of \cite{kovtun} in Sections \ref{shear} and \ref{num}. 

It is appropriate at this point to emphasize that our approach is purely phenomenological. It is not our intention to give a precise value of the correction to the microscopic shear viscosity arising from second order terms, but rather to provide a reasonable upper bound for it under realistic conditions for the QGP. We note that the correction that we obtain, as well as that obtained in \cite{kovtun}, is inversely proportional to $\eta/s$, and therefore diverges for vanishing $\eta/s$. As pointed out recently by Torrieri \cite{torrieri} in the context of ideal quantum hydrodynamics, the cure for this divergence may lie in quantum corrections that set in for small $\eta/s$ and therefore ``correct the correction''. Although this is an important issue with strong implications for the study of matter created in heavy ion collisions and surely deserves further attention, here we shall not dwelve further into it since we focus on the correction to $\eta/s$ arising at finite values of this ratio.

The paper is organized as follows. In Section \ref{fluc} we describe the theoretical setup and obtain the equations for the ultrasoft and soft modes. In Section \ref{shear} we compute the correction to the shear viscosity arising from second order soft terms, and then provide numerical estimates for a weakly and a strongly coupled theory, in particular focusing on the QGP created in heavy ion collisions. Finally, we conclude with a brief summary and outlook in Section \ref{summary}. Details on the calculation of the correction to $\eta/s$ are given in Appendix \ref{app}, while a description of the procedure used to compute the frequency that separates ultrasoft and soft modes is given in Appendix \ref{omin}. 

\section{Fluctuating conformal hydrodynamics}
\label{fluc}

\subsection{Soft and ultrasoft hydrodynamic modes}
We consider a conformal fluid in flat spacetime with signature $(+,-,-,-)$. The comoving time and space derivatives are $D=u_\mu \partial^\mu$ and $\nabla^\mu=\Delta^{\mu\nu}\partial_\nu$, respectively. The hydrodynamic equations at second order in velocity gradients read
\begin{equation}
\begin{split}
\frac{4}{3}\rho D u^\mu - \frac{1}{3}\nabla^\mu \rho + \eta \Delta_\alpha^\mu \partial_\beta \sigma^{\alpha \beta} &=f^\mu\\
D\rho +\frac{4}{3} \rho \nabla_\mu u^\mu - \eta \sigma^{\mu\nu}\sigma_{\mu\nu} &= f
\end{split}
\label{hydroeq}
\end{equation}
where we have made the approximation $\Pi^{\mu\nu}=\eta \sigma^{\mu\nu}=\eta \nabla^{<\mu} u^{\nu>}$ which is enough for our purposes here. The brackets $<>$ around indices denote the symmetric and traceless projection orthogonal to $u^\mu$, while round brackets denote symmetrization. $\eta$ is the microscopic shear viscosity as derived e.g. by Kubo's formula, and $(f,f^\mu)$ are stochastic noises. For recent reviews on relativistic viscous fluid dynamics as applied to heavy ion collisions see Refs. \cite{rev1,rev2,rev3}.

Next we divide the hydrodynamic fields into ``ultrasoft'' and ``soft'' parts, 
\begin{equation}
\rho = \rho_< + \rho_> ~~ \textrm{and}~~ u^\mu =  u^\mu_< + u^\mu_> 
\end{equation}
On top of this separation into ultrasoft and soft sectors, we perfom an expansion in deviations from equilibrium characterized by a small parameter, $\epsilon$, staying at linear order, thus
\begin{equation}
\begin{split}
\rho_< &= \rho_{0,<} + \epsilon \rho_{1,<} + \ldots \\
 \rho_> &= \rho_{0,>} + \epsilon \rho_{1,>} + \ldots \\
u^\mu_< &= u_{0,<}^\mu + \epsilon u_{1,<}^\mu + \ldots \\
u^\mu_>  &= u_{0,>}^\mu + \epsilon u_{1,>}^\mu + \ldots 
\end{split}
\end{equation}
In this way $(\rho_{0,<}+\rho_{0,>},u_{0,<}^\mu+u_{0,>}^\mu)$ correspond to thermal equilibrium. For notational simplicity is it convenient to define 
\begin{equation}
 \begin{split}
\rho_0 &\equiv \rho_{0,<} \qquad , \qquad \rho_1 \equiv \rho_{1,<} \\ 
\delta \rho_0 &\equiv \rho_{0,>} \qquad , \qquad \delta \rho_1 \equiv  \rho_{1,>} \\ 
V^\mu &\equiv u_{0,<}^\mu \qquad , \qquad W^\mu \equiv u_{1,<} \\ 
s^\mu &\equiv  u_{0,>}^\mu  \qquad , \qquad t^\mu \equiv u_{1,>}^\mu   
 \end{split}
\end{equation}
 It will be shown in the next section that $V^\mu = (V^0,0,0,0)$, with $V^0$ close to but not exactly unity, and that $V_\mu s^\mu = V_\mu W^\mu = 0$, which imply that $s^\mu = (0,\vec{s})$ and $W^\mu = (0,\vec{W})$ in the local rest frame (LRF). We note that $V_\mu t^\mu \neq 0$.  Additional constrains arising from the normalization of $u^\mu$ are discussed in the next section. 


Note that our treatment is purely hydrodynamic, i.e. assuming momenta to be small, so no attempt is made to actually derive the noise as coming from the {\it hard} sector of a microscopic field which should be described by transport equations such as Boltzmann equation \cite{libro,Bod98,Bod99,ArSoYa99a,ArSoYa99b,LitMan02}. So, the ultrasoft and soft modes of the hydrodynamic fields are described by the conservation equations of fluid dynamics, while the effect of the hard modes on the hydrodynamic evolution is included through the noises $(f,f^\mu)$. This is not a limitation to our purpose here, since we intend to determine the correction to $\eta$ coming from long--lived {\it hydrodynamic} modes (shear and sound waves).

A reasonable estimate for the value of the momentum separating soft and hard modes will be given in Section \ref{shear}. This estimate is essentially based on the requirement that second order terms in derivatives must be smaller than first order terms, i.e. that the gradient expansion does not break down. The value of momentum that separates ultrasoft from soft modes can be computed by requiring that the equations of the soft modes be linearizable (see, e.g., \cite{libro,blaizot}). This is done in Appendix \ref{omin}. Due to the phenomenological nature of our approach, and taking into account that we intend to provide an upper bound to the correction to $\eta/s$, we also compute this correction for different values of the soft--ultrasoft separation frequency. We will come back to this point in Section \ref{shear}. For the moment, let us leave this separation implicit. 

In what follows we will use Latin indices $(i,j,k)$ to denote spatial coordinates, i.e. $s^\mu=(0,s^i)$ with $i=1,2,3$. At zeroth order in $\epsilon$ the conservation equations for the ultrasoft modes read
\begin{equation}
\begin{split}
& \left\langle \delta D_0 \delta \rho_0 \right\rangle + \frac{4}{3}\left\langle \delta \rho_0 \nabla_{(0)i} s^i \right\rangle 
-\frac{4}{3}\left\langle \rho_0 V_{0}s_{i}\partial^0 s^i \right\rangle - \eta \left\langle s^{ij} s_{ij} \right\rangle = 0 \\
& \frac{4}{3} \left\langle \rho_0 \delta D_0 s^\mu \right\rangle + \frac{4}{3} \left\langle \delta \rho_0 D_0 s^\mu \right\rangle 
+ \frac{2}{3}\left\langle V^{(\mu}s^{\phi)} \partial_\phi \delta \rho_0 \right\rangle - \eta \left\langle V^{\mu}s^{i}\partial^j s_{ij} \right\rangle = 0
\end{split}
\label{1s}
\end{equation}
where $\left\langle \cdot \right\rangle$ is a thermal average. 
The zeroth order linearized equations for the soft modes are 
\begin{equation}
\begin{split}
& D_0 \delta \rho_0 + \frac{4}{3}\rho_0 \nabla_{(0)i}s^i = f \\
& \frac{4}{3}\rho_0 D_0 s^i -\frac{1}{3} \nabla_{(0)}^i \delta \rho_0 + \eta \Delta^i_{(0)j} \partial_k s^{jk} = f^i ~.
\end{split}
\label{1h}
\end{equation}
Note that the noises are determined by the equilibrium state, as dictated by the fluctuation--dissipation theorem. Their role is to enforce thermal equilibrium at zeroth order in $\epsilon$, constraining the expression for the thermal correlator of velocity fluctuations which will be given in Section \ref{shear}.

At first order we get for the ultrasoft modes 
\begin{equation}
\begin{split}
& \left\langle \delta D_0 \delta \rho_1 \right\rangle + \left\langle \delta D_1 \delta \rho_0 \right\rangle + 
\frac{4}{3}\rho_0 \nabla_{(0)i}W^i + \frac{4}{3}\left\langle \delta \rho_1 \nabla_{(0)i}s^i \right\rangle \\
&- \frac{8}{3}\left\langle \rho_0 V_{(\mu} s_{\phi)}\partial^\phi t^\mu\right\rangle + 
\frac{4}{3}\left\langle \delta \rho_0 \nabla_{(0)\mu}t^\mu \right\rangle 
- \frac{4}{3}\left\langle \delta \rho_0 V_{0}s_{i}\partial^0 W^i \right\rangle \\
&- \frac{4}{3}\left\langle \delta \rho_0 V_{0}W_{i}\partial^0 s^i \right\rangle 
- \frac{8}{3}\left\langle \delta \rho_0 W_{(i}s_{j)}\partial^i s^j \right\rangle 
- \frac{4}{3}\left\langle \rho_0 V_{0}t_{i}\partial^0 s^i \right\rangle -2\eta \left\langle s_{ij}t^{ij} \right\rangle = 0
\end{split}
\label{2s1}
\end{equation}
and 
\begin{equation}
\begin{split}
& \frac{4}{3}\rho_0 D_0 W^\mu + \frac{4}{3}\left\langle \rho_0 \delta D_0 t^\mu \right\rangle 
+ \frac{4}{3}\left\langle \rho_0 \delta D_1 s^\mu \right\rangle \\
&+ \frac{4}{3}\left\langle \delta \rho_0 D_0 t^\mu \right\rangle 
+ \frac{4}{3}\left\langle \delta \rho_0 \delta D_0 W^\mu \right\rangle 
+ \frac{4}{3}\left\langle \delta \rho_0 D_1 s^\mu \right\rangle \\
&+ \frac{4}{3}\left\langle \delta\rho_1 D_0 s^\mu \right\rangle  
+ \frac{2}{3}\left\langle V^{(\mu}s^{\phi)}\partial_\phi \delta \rho_1 \right\rangle 
+ \frac{2}{3}\left\langle W^{(\mu}s^{\phi)}\partial_\phi \delta \rho_0 \right\rangle \\
& + \frac{2}{3}\left\langle V^{(\mu}t^{\phi)}\partial_\phi \delta \rho_0 \right\rangle  
+ \eta C^\mu = 0
\end{split}
\label{2s2}
\end{equation}
with 
\begin{equation}
\begin{split}
C^\mu &= \Delta_{(0)\alpha}^\mu \partial_\beta W^{\alpha\beta} 
-2\left\langle \Delta_{(0)\alpha}^\mu \partial_\beta (W^{<(\alpha}s^{\phi)}\partial_\phi s^{\beta>}) \right\rangle -2g_{\gamma\alpha}\left\langle V^{(\mu}s^{\gamma)} \partial_\beta t^{\alpha\beta} \right\rangle \\
& -2g_{\gamma\alpha}\left\langle W^{(\mu}s^{\gamma)} \partial_\beta s^{\alpha\beta} \right\rangle 
 -2g_{\gamma\alpha}\left\langle V^{(\mu}t^{\gamma)} \partial_\beta s^{\alpha\beta} 
 \right\rangle .
\end{split}
\end{equation}

For the soft modes we get 
\begin{equation}
\begin{split}
& D_0 \delta \rho_1 + D_1 \delta \rho_0 - \frac{4}{3}\rho_0 V_{0}W_{i}\partial^0 s^i \\
& + \frac{4}{3}\rho_0 \nabla_{(0)\mu}t^\mu + \frac{4}{3}\delta\rho_0 \nabla_{(0)i}W^i 
- 2\eta s_{ij}W^{ij} = 0
\end{split}
\label{2h1}
\end{equation}
and 
\begin{equation}
\frac{4}{3}\rho_0 D_0 t^\mu + \frac{4}{3}\delta \rho_0 D_0 W^\mu - \frac{1}{3}\nabla_{(0)}^\mu \delta \rho_1 
+ \frac{2}{3}V^{(\mu}W^{\phi)}\partial_\phi \delta \rho_0 + \eta E^\mu = 0
\label{2h2}
\end{equation}
with
\begin{equation}
E^\mu = \Delta^\mu_{(0)\alpha}\partial_\beta t^{\alpha \beta} -2 g_{\gamma\alpha}V^{(\mu}s^{\gamma)}\partial_\beta W^{\alpha\beta} 
-2 g_{\gamma\alpha}V^{(\mu}W^{\gamma)}\partial_\beta s^{\alpha\beta} ~~.
\label{2h2e}
\end{equation}

In deriving Eqs. (\ref{1s})--(\ref{2h2e}) we have used that 
\begin{equation}
\nabla^\mu = \nabla^\mu_{(0)} - 2 V^{(\mu}s^{\phi)}\partial_\phi - 2 \epsilon V^{(\mu}W^{\phi)}\partial_\phi 
- 2 \epsilon W^{(\mu}s^{\phi)}\partial_\phi - 2 \epsilon V^{(\mu}t^{\phi)}\partial_\phi 
\label{nabla}
\end{equation}
where $\nabla^\mu_{(0)}=(g^{\mu\nu}-V^\mu V^\nu)\partial_\nu$ and $A^{(\mu\nu)}=(A^{\mu\nu}+A^{\nu\mu})/2$, 
\begin{equation}
D = D_0 + \delta D_0 + \epsilon D_1 + \epsilon \delta D_1
\end{equation}
where $D_0 = V^\mu \partial_\mu$, $\delta D_0 = s^\mu \partial_\mu$, $D_1 = W^\mu \partial_\mu$ and $\delta D_1 = t^\mu \partial_\mu$, and put 
\begin{equation}
\begin{split}
s^{\mu\nu} &\equiv \nabla^{<\mu}_{(0)}s^{\nu>} \\
t^{\mu\nu} &\equiv \nabla^{<\mu}_{(0)}t^{\nu>} \\
W^{\mu\nu} &\equiv \nabla^{<\mu}_{(0)}W^{\nu>} 
\end{split}
\end{equation}
which by definition are traceless and orthogonal to $V_\mu$. Note that we have neglected terms $O(\delta^3)$ or higher in the thermal averages. Moreover, we have set $\rho_1 = 0$, which is a reasonable approximation since, as it will be seen later, the term we are interested in is already linear in $\epsilon$. Note that, since we stay at first order and $O(\delta^3)$, those terms containing $V^{<\mu}$ that would appear in Eqs. (\ref{1s})--(\ref{2h2e}) actually  vanish because in these terms we can put $\Delta^{\mu\nu}=\Delta^{\mu\nu}_{(0)}$ (see Eq. (\ref{nabla})).

\subsection{Constraints on four velocity}
We will now discuss the normalization of the fluid velocity. For consistency, the velocity should be normalized both in the mean and at linear order in fluctuations, i.e. $u^\mu u_\mu = 1$, because it is $u^\mu$ who satisfies the conservation equations (\ref{hydroeq}). We also require that $(V^\mu+s^\mu)^2 = 1$ which implies 
\begin{equation}
\begin{split}
V^\mu V_\mu + \left\langle s^\mu s_\mu \right\rangle &= 1 ~~ \textrm{and} \\
V^\mu s_\mu &= 0
\end{split}
\label{norm1}
\end{equation}
so at first order we get 
\begin{equation}
\begin{split}
V^\mu W_\mu = 0   ~~ \textrm{and} \\
V^\mu t_\mu + W^\mu s_\mu &= 0 ~.
\end{split}
\label{norm2}
\end{equation}

The transversality and tracelessness of $\sigma^{\mu\nu}$ must be satisfied both in the mean and at linear order in fluctuations. The constraints coming from the transversality and tracelessness of $\sigma^{\mu\nu}$ then read
\begin{equation}
\begin{split}
\left\langle s^\mu s_{\mu\nu}  \right\rangle &= 0 \\
\left\langle  V^{(\mu}s^{\phi)}\partial_\phi s_\mu  \right\rangle &= 0 \\
\left\langle t^\mu s_{\mu\nu} \right\rangle &= 0 \\
\left\langle  V^{(\mu}s^{\phi)}\partial_\phi t_\mu \right\rangle 
+\left\langle  W^{(\mu}s^{\phi)}\partial_\phi s_\mu \right\rangle 
+ \left\langle  V^{(\mu}t^{\phi)}\partial_\phi s_\mu \right\rangle &= 0 \\
s^\mu W_{\mu\nu} + W^\mu s_{\mu\nu} &= 0 \\
V^{(\mu}s^{\phi)}\partial_\phi W_\mu + V^{(\mu}W^{\phi)}\partial_\phi s_\mu &= 0
\end{split}
\label{cons}
\end{equation}
These relations were used in deriving Eqs. (\ref{1s})--(\ref{2h2e}) and will be used in what follows. 

\section{Shear viscosity induced by the soft modes}
\label{shear}

The idea is now to solve the soft mode equations to compute the induced viscosity arising in the ultrasoft equations. We will solve the soft mode equations in Fourier space. For any given quantity $R(x^\mu)$ we have 
\begin{equation}
\tilde{R} (k^\mu) = \mathfrak{F}[R(x)](k^\mu) = \int R(x^\mu) e^{ik_\mu x^\mu} ~d^4 x
\end{equation}
where $\mathfrak{F}[g(x)](k_\mu)\equiv \tilde{g}(k_\mu)$ denotes the Fourier transform. We split the wavevector $k^\mu = \omega n^\mu + p^\mu$, where $n^\mu = V^\mu/|V^2|$. Since $V^\mu=(V^0,0)$ in the LRF and $V^\mu s_\mu = 0$ we have $s^\mu=(0,\vec{s})$, and similarly for $W^\mu$ (but not for $t^\mu$). Note that $p^\mu = (0,p^i)$ in the LRF.

The zeroth order equations then read
\begin{equation}
\begin{split}
\omega \delta \tilde{\rho}_0 + \frac{4}{3}\rho_0 p_j \tilde{s}^j &= i \tilde{f} ~~~ \textrm{and} \\
\frac{4}{3}\rho_0 \omega \tilde{s}^i - \frac{1}{3}p^i \delta \tilde{\rho}_0 - \frac{i\eta}{2} \bigg[\frac{1}{3}p^i (p_j\tilde{s}^j) + 
\tilde{s}^i p^j p_j \bigg] &= i \tilde{f}^i ,~~~ i = 1,2,3
\end{split}
\label{hard0}
\end{equation}
Without loss of generality we can set $p_2=p_3=0$. The solution is ($p \equiv p_1$)
\begin{equation}
\begin{split}
\delta \tilde{\rho}_0 &= 3i\frac{i\eta\tilde{f}p^2 + 2\tilde{f}\rho_0 \omega + 2\tilde{f}_1 \rho_0 p}{A}\\
\tilde{s}_1 &= 3i\frac{3\tilde{f}_1 \omega +\tilde{f}p}{2A} \\
\tilde{s}_2 &= \frac{6\tilde{f}_2}{B} \\
\tilde{s}_3 &= \frac{6\tilde{f}_3}{B} 
\end{split}
\end{equation}
where 
\begin{equation}
\begin{split}
A &= 3i\eta \omega p^2 + 6\rho_0(\omega^2 -\frac{1}{3} p^2) ~~~ \textrm{and} \\
B &= 3\eta p^2 - 8 i \rho_0 \omega 
\end{split}
\end{equation}

The first order soft mode equations are
\begin{equation}
\begin{split}
\omega \delta \tilde{\rho}_1 + \frac{4}{3}\rho_0 p_j \tilde{t}^j &= -i G[W^\sigma] \\
-\frac{1}{3}p^i \delta \tilde{\rho}_1 + \frac{4}{3}\rho_0 \omega \tilde{t}^i -\frac{i\eta}{2}\bigg[ \frac{1}{3}p^i (p_j\tilde{t}^j) + 
\tilde{t}^i p^j p_j\bigg]&= -iH^i[W^\sigma] \\
\frac{4}{3}\rho_0 \omega \tilde{t}^0 &= iH^0[W^\sigma]
\end{split}
\label{hard1}
\end{equation}
where
\begin{equation}
\begin{split}
G[W^\mu] &= \int e^{ik_\sigma x^\sigma} \bigg[W^\mu \partial_\mu \delta \rho_0 
-\frac{8}{3}\rho_0 V_{(\mu}W_{\phi)}\partial^\phi s^\mu \\
&+ \frac{4}{3}\delta \rho_0 \nabla_{(0)\mu}W^\mu - 2\eta s_{\mu\nu}W^{\mu\nu} \bigg]~ d^4x\\
H^\mu [W^\mu] &= \int  e^{ik_\sigma x^\sigma} \bigg[ 
\frac{4}{3}\delta \rho_0 D_0 W^\mu + \frac{2}{3}V^{(\mu}W^{\phi)}\partial_\phi \delta \rho_0 \\
&- 2\eta g_{\gamma\alpha}V^{(\mu}s^{\gamma)}\partial_\beta W^{\alpha\beta} 
 - 2\eta g_{\gamma\alpha}V^{(\mu}W^{\gamma)}\partial_\beta s^{\alpha\beta} \bigg] ~ d^4x 
\end{split}
\label{GH}
\end{equation}
whose solution is  
\begin{equation}
\begin{split}
\delta \tilde{\rho}_1 &= 3i\frac{i\eta G p^2 + 2G\rho_0 \omega + 2 H_1 \rho_0 p}{A}\\
\tilde{t}_1 &= 3i\frac{3 H_1 \omega +G p}{2A} \\
\tilde{t}_2 &= \frac{6H_2}{B} \\
\tilde{t}_3 &= \frac{6H_3}{B} \\
\tilde{t}_0 &= i\frac{3}{4\rho_0 \omega}H_0
\end{split}
\label{solt}
\end{equation}

We are interested in those terms appearing in Eq. (\ref{2h2}) containing two (orthogonal to $V^\mu$) derivatives of $W^\gamma$. Moreover, since the bare term $\Delta_{(0)\alpha}^\mu \partial_\beta W^{\alpha\beta}$ is orthogonal  to $V_\mu$, the induced terms that provide the correction to $\eta$ must be orthogonal to $V_\mu$ as well. The only term in $H^\mu$ proportional to $\partial_\beta W^{\alpha\beta}$ is 
\begin{equation}
 - 2\eta g_{\gamma\alpha}V^{(\mu} \int  e^{ik_\sigma x^\sigma} s^{\gamma)}\partial_\beta W^{\alpha\beta} ~ d^4x = 
- \eta V^\mu \int  e^{ik_\sigma x^\sigma}  s_\alpha \partial_\beta W^{\alpha\beta} ~ d^4x
\end{equation}
since $V_\alpha \partial_\beta W^{\alpha\beta} = 0$. Therefore, this term in $H^\mu$ is proportional to $V^\mu$. So, those terms in Eq. (\ref{2h2}) which upon replacement of $t^\mu$ by its expression given in Eq. (\ref{solt}) contain two derivatives of $W^\mu$  are not orthogonal to $V_\mu$, but proportional to it instead. Examples of such term are $\left\langle 4\delta \rho_0 D_0 t^\mu /3\right\rangle$ and 
$-2\eta g_{\gamma\alpha}\left\langle V^{(\mu}t^{\gamma)}\partial_\beta s^{\alpha\beta} \right\rangle$. Note that the latter term vanishes identically because $V_\alpha \partial_\beta s^{\alpha\beta} = 0$.  
On the other hand, a term such as $\left\langle 4\delta \rho_1 D_0 s^\mu \right\rangle$ does not contribute to the correction to $\eta$, since $\delta \rho_1$ depends on $H_1$, whose term proportional to $\partial_\beta W^{\alpha\beta}$ vanishes (it is proportional to $V_1=0$). 

Therefore, the only term satisfying all the requirements mentioned  above is $\left\langle 4 \rho_0 \delta D_1 s^\mu /3\right\rangle$. This term is orthogonal to $V_\mu$ and is second order in spatial derivatives of $W^\mu$. 
The correction to $\eta$, which we denote as $\eta_c$, comes from the coefficient multiplying $ \partial_\beta W^{\alpha\beta}$ in this term. We focus on the $\mu=1$ component, i.e. we only consider the $x$ component of the momentum conservation equation. To be definite, we take $\alpha=2$ and  $\beta=1$, which means that we are considering the correction to $\eta$ that comes from the term $\partial_x \nabla_{(0)}^y W^x$.  Note that from $\delta D_1 = t^\alpha D_\alpha$ appearing in the term $\left\langle 4 \rho_0 \delta D_1 s^\mu /3\right\rangle$ we only need  $t^0$, because this component is the only one containing $\partial_\beta W^{\alpha\beta}$ -- see Eqs. (\ref{GH}) and (\ref{solt}). Taking into account these considerations,  the term we are interested in becomes (for simplicity we use $x$ instead of $x^\mu$)
\begin{equation}
A(x) \equiv \frac{4\rho_0}{3}\left\langle  t^0(x) D_0(x) s^1(x) \right\rangle 
\label{uno}
\end{equation}
together with the equation
\begin{equation}
P(x)t^0(x) = \eta s_2(x) dW(x)
\label{dos}
\end{equation}
where we have defined the operator $P(x)=4\rho_0 D_0/3$, and set $dW\equiv \partial_x W^{yx}$. For simplicity, we have approximated $V^0=1$ as well, which implies neglecting $\left\langle s_i s^i\right\rangle$ in the normalization of $u^\mu$, see Eq. (\ref{norm1}). The term given in Eq. (\ref{uno}), from which the correction to the shear viscosity $\eta_c$ arises, is quadratic in fluctuations and second order in spatial gradients (because $t^0$ is proportional to $dW$ which is second order). 
To obtain $\eta_c$ we must solve Eq. (\ref{dos}) and plug it into Eq. (\ref{uno}). Details of this procedure are given in Appendix \ref{app}. The expression for $\eta_c$ that results is given in Eqs. (\ref{etac11})-(\ref{etac3}) and Eq. (\ref{etac-sound}). 

In order to give concrete numerical estimates for $\eta_c/s$ in the next section, it will prove convenient to express $\eta_c$ in terms of dimensionless quantities. The relevant parameters appearing in the expression for $\eta_c$ that is given in Eqs. (\ref{etac11})-(\ref{etac3}) and Eq. (\ref{etac-sound}) are $(\omega_{max},p_{max})$ -- the maximum momentum for the soft modes, beyond which the gradient expansion becomes invalid --, $\omega_{min}$ (the frequency separating ultrasoft and soft modes), and $\gamma=\eta/(sT)$, where $s$ is the entropy density. 
We therefore define the two dimensionless quantities 
\begin{equation}
\chi=\gamma p_{max}
\end{equation}
and 
\begin{equation}
\xi = \frac{\omega_{min}}{\omega_{max}}
\end{equation}
 
The value of  $p_{max}$ can be roughly estimated as $p_{max} \sim 1/(2\tau_\pi)$, where $\tau_\pi$ is the second order coefficient representing the relaxation time of the full shear tensor $\Pi^{\mu\nu}$ towards its Navier-Stokes value $\Pi^{\mu\nu} = \eta \sigma^{\mu\nu}$. The value of $\tau_\pi$ has been computed in strongly coupled $\mathcal{N}=4$ super Yang--Mills theory \cite{hyd1,hyd2,hyd3}, resulting in $\tau_\pi \sim 2.6 \eta/(sT) = 2.6\gamma$, and in a weakly coupled theory from kinetic theory \cite{ttr11}, resulting in $\tau_\pi \sim 5\gamma$. The trend is that $\tau_\pi/\gamma$ increases as the coupling decreases. Here, we shall use as a reasonable estimate a value of $\tau_\pi = 3\gamma$ to model the strongly coupled QGP, and a value of $\tau_\pi = 6\gamma$ to model a weakly coupled theory. It is important to note that, as it has been recently recognized \cite{deni}, the true relaxation time appearing in the equation of motion of dissipative currents is given by the first pole of the retarded Green function.  It turns out that the location of this pole can not be found, in general (and particularly for a strongly coupled theory), from a gradient expansion. In other words, the equation of motion for the dissipative current is
not a relaxation-type equation -- see \cite{deni} for details. Nevertheless, in this paper we will use $\tau_\pi \propto \gamma$ as computed from the second order gradient expansion \cite{hyd1,hyd2,hyd3,ttr11}, both for weakly and strongly coupled settings.  

We note that by taking the value $p_{max} = 1/(2\tau_\pi)$, the validity of the gradient expansion, i.e. the fact that higher order terms are smaller than lower order ones, is ensured, since for this to happen one must have $p_{max} < 1/\tau_\pi$ (of course, second order fluid dynamics becomes more and more reliable the smaller $p_{max}\tau_\pi$ is) -- see \cite{kovtun}. This condition also ensures that hydrodynamic waves are actually long--lived excitations of the fluid and thus well defined. As an estimative value for $\omega_{max}$, we shall take $\omega_{max} \sim 1/(2\tau_\pi)$ as well. In Section \ref{num} we will consider other values of $p_{max}$ and analyze the dependence of our results on these values. 

It is useful to note that since $\tau_\pi \sim \eta/(sT)$, a term containing $p^2_{max}T$ in $\eta_c$ is proportional to  $T^3 (\eta/s)^{-2}$. If we put $s=aT^3$ (estimates of the value of $a$ for strongly and weakly coupled theories are given in the next section) we get that  
\begin{equation}
 p^2_{max} T= \frac{s \chi^2}{a(\eta/s)^2}
\label{inv}
\end{equation}

To simplify the resulting expression for $\eta_c/s$, let us define the following quantities 
\begin{equation}
\begin{split}
L_{max}&=\frac{1+\sqrt{2\chi}+\chi}{1-\sqrt{2\chi}+\chi}  \qquad \textrm{and} \qquad  L_{min} = \frac{\xi+\sqrt{2\xi\chi}+\chi}{\xi-\sqrt{2\xi\chi}+\chi} \\
M &= \frac{\chi(1+\xi)}{\xi-\chi^2} \\
N_{max} &= \frac{\sqrt{2\chi}}{1-\chi} \qquad \textrm{and} \qquad N_{min} = \frac{\sqrt{2\xi\chi}}{\xi-\chi} \\
K &= \frac{\chi^2}{2\pi^3 a (\eta/s)^2}
\end{split}
\end{equation}
We then have (see  Eqs. (\ref{etac11})-(\ref{etac3}))
\begin{equation}
 \begin{split}
\frac{\eta_c}{s} &= K \bigg[  \frac{4}{3}(1-\xi) -\frac{1}{3}\pi \chi-\frac{2\chi}{3}\textrm{arctan}~M  -\frac{1}{3\sqrt{2\chi}}\ln L_{max} \\
&+ \frac{2\xi^{3/2}}{3\sqrt{2\chi}} \ln L_{min} - \frac{2}{3\sqrt{2\chi}}\textrm{arctan}~N_{max} 
+ \frac{4\xi^{3/2}}{3\sqrt{2\chi}} \textrm{arctan}~N_{min} \bigg] 
 \end{split}
\label{final}
\end{equation}

The expression for the correction to the shear viscosity given in Eq. (\ref{final}) is our main result. 
We have therefore computed the correction to $\eta$ that comes from the action of the soft modes on the evolution of the ultrasoft modes, as is reflected in the second order term $\partial_x \nabla_{(0)}^y W^x$ of the fluid dynamic equations.  

At this point, it is important to stress that out treatment is, although similar in spirit, otherwise quite different from the one put forward in Ref. \cite{kovtun} in two main points. First, the developments of \cite{kovtun} are based on Kubo relations for $\eta$ and $\tau_\pi$ in terms of the retarded correlation function for two $T^{xy}$ stress tensor operators, while ours rely on the explicit separation of hydrodynamic modes into ultrasoft and softs components that is done on top of an expansion in deviations from thermal equilibrium. Second, the authors of \cite{kovtun} focus on the term $(\rho_0 + P_0)u^\mu u^\nu$ of the stress energy tensor (the background velocity is zero), which does not contain gradients, while we focus on an induced term in the conservation equations that is second order in (background) velocity gradients (therefore the term comes from a first order term in the stress energy tensor). The contribution to $\eta_c$ of the term that we consider is suppressed with respect to the zeroth order term by extra powers of $\chi = p_{max}\gamma$. The zeroth order term $\sim u^2$ leads to a correction to $\eta/s$ that is of $O(\chi)$, which in our notation reads [see Eq. (4.2) of \cite{kovtun}]:
\begin{equation}
\frac{\eta_c}{s} = \frac{17 \chi}{120\pi^2 a (\eta/s)^2}
\end{equation}
As it can be seen from Eq. (\ref{final}),  the contribution to $\eta_c/s$ found here is $O(\chi^2)$ or higher.
In the next section we will analyze the behavior of $\eta_c/s$ with $\eta/s$ for strongly and weakly coupled theories and compare the obtained results with those of \cite{kovtun}.

\subsection{Numerical estimates for weakly and strongly coupled theories}
\label{num}

We will now provide some estimates for the strongly coupled QGP as well as for an illustrative weakly coupled theory. We shall take the entropy density to be $s \sim a T^3$, with $a=N_c^2$ ($N_c$ is the number of colors) for a weakly coupled theory and $a = 10$ as a typical value for the QGP in the temperature range $200$--$300$ MeV (see \cite{bor,baza}). With respect to the second order transport coefficient $\tau_\pi$, for the QGP we will take $\tau_\pi \sim 3\gamma$  \cite{hyd1}, which implies that $\chi = p_{max}\gamma = 1/6$ (if $p_{max}=1/(2\tau_\pi)$), while as a typical value for a weak coupling theory we will take $\chi = 1/12$ \cite{ttr11}. Recall also that we take $\omega_{max}=p_{max}$. 

In the following we will focus on values of $\eta/s$ in the range of those values that can be extracted by matching fluid dynamic simulations to elliptic flow data -- see \cite{ext1,ext2,ext3,ext4,PRC-E}. In particular, although we will take the values of $\eta/s$ to be in the range $0.06$--$0.16$, we will pay special attention to the comparison between the cases $\eta/s=0.08$ and $\eta/s=0.16$. The reason for this choice is that $\eta/s=0.08$ is close to the KSS lower bound obtained from the AdS/CFT correspondence \cite{KSS},  while $\eta/s = 0.16$ is close to the maximum value of $\eta/s$ that is favored by fitting fluid dynamic simulations to elliptic flow data \cite{ext1,ext2,ext3,ext4,PRC-E}. 

To estimate $\eta_c/s$ we still need to provide values of $\omega_{min}$. A natural request to our approach is that $\omega_{min}$ must be such that the equations for the soft modes can actually be linearized. This condition is analyzed in detail in Appendix \ref{omin}. Here we will just quote the values obtained from this analysis. For a strongly coupled theory with $\chi=1/6$ we get from Eq. (\ref{xiapp2}) that $\xi = \omega_{min}/\omega_{max}=0.88$ for $\eta/s=0.08$ and $\xi=0.72$ for $\eta/s=0.16$. For a weakly coupled theory with $\chi=1/12$  we get 
$\xi =0.78$ for $\eta/s=0.08$ and $\xi=0.56$ for $\eta/s=0.16$.

\begin{center}
\begin{figure}
\scalebox{0.8}{\includegraphics{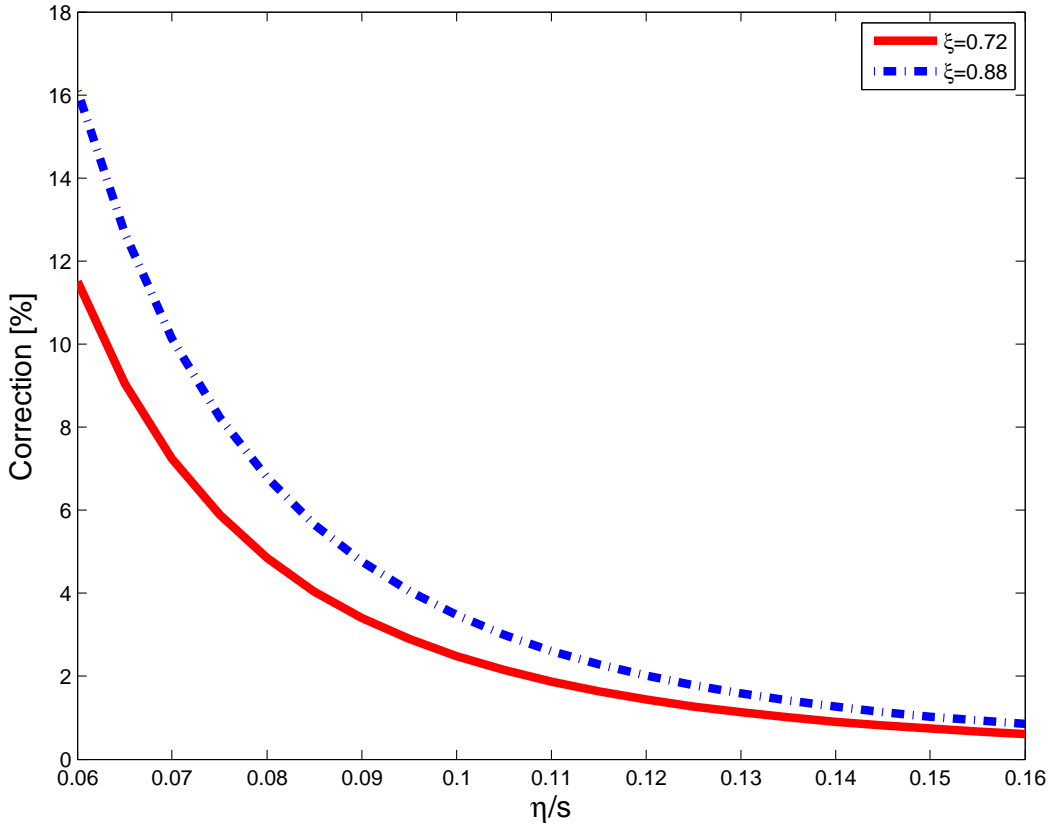}}
\caption{(Color online) Percentual correction  $(\eta_c/s)\times 100/(\eta/s)$ as a function of $\eta/s$ for $\xi = \omega_{min}/\omega_{max}=0.88$ and $\xi = 0.72$. The value of $\chi=p_{max}\gamma$ is $1/6$, corresponding to a strongly coupled theory.}
\label{f1}
\end{figure}
\end{center}

With the estimates given above we can compute $\eta_c/s$ as a function of $\eta/s$ in the range of values of relevance to heavy ion collisions. Figure \ref{f1} shows the percentual correction $(\eta_c/s)\times 100/(\eta/s)$ as a function of $\eta/s$, for $\chi=p_{max}\gamma = 1/6$ (corresponding to the strongly coupled regime). The results are shown for two values of $\xi=\omega_{min}/\omega_{max}=0.88$ and $\xi=0.72$ corresponding to the values of $\omega_{min}$ computed in Appendix \ref{omin} for $\eta/s = 0.08$ and $\eta/s = 0.16$ (see Eq. (\ref{xiapp2})). It is seen that for $\eta/s \sim 0.08$ the correction amounts to roughly $8 \%$ of $\eta/s$. Due to the inverse proportionality of $\eta_c/s$ on the square of the microscopic ratio, the correction drops to zero quite fast. For $\eta/c = 0.16$, the correction is less than $1\%$.

\begin{center}
\begin{figure}
\scalebox{0.8}{\includegraphics{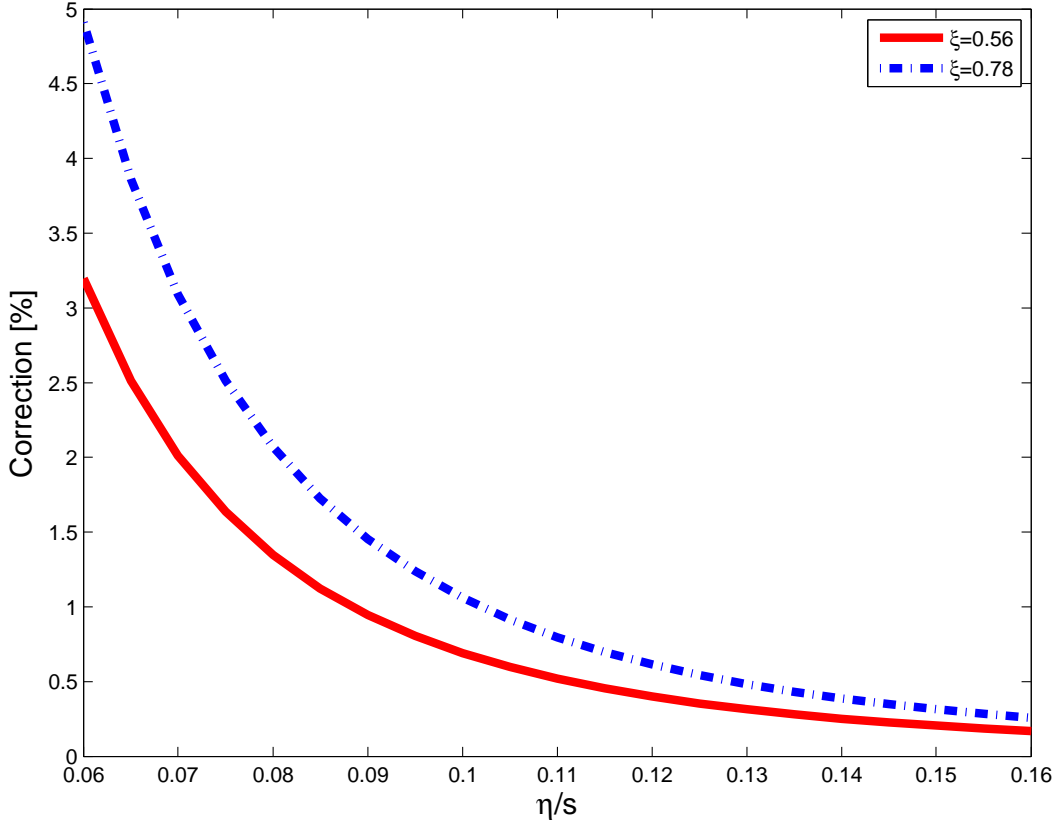}}
\caption{(Color online) Percentual correction  $(\eta_c/s)\times 100/(\eta/s)$ as a function of $\eta/s$ for $\xi = \omega_{min}/\omega_{max}=0.78$ and $\xi = 0.56$. The value of $\chi=p_{max}\gamma$ is $1/12$, corresponding to a weakly coupled theory.}
\label{f2}
\end{figure}
\end{center}

Going over to a weakly coupled theory, Figure \ref{f2} shows, as an illustrative case, the percentual correction $(\eta_c/s)\times 100/(\eta/s)$ as a function of $\eta/s$ for $\chi=p_{max}\gamma=1/12$ and $a=9$ (corresponding to $N_c=3$ colors). It is seen that the correction $\eta_c/s$ is much smaller than the one obtained in a strongly coupled setting. For $\eta/s=0.08$, the correction is 
$\sim 2\%$, whereas for $\eta/s =0.16$ it amounts to less than half percent.

\begin{center}
\begin{figure}
\scalebox{0.8}{\includegraphics{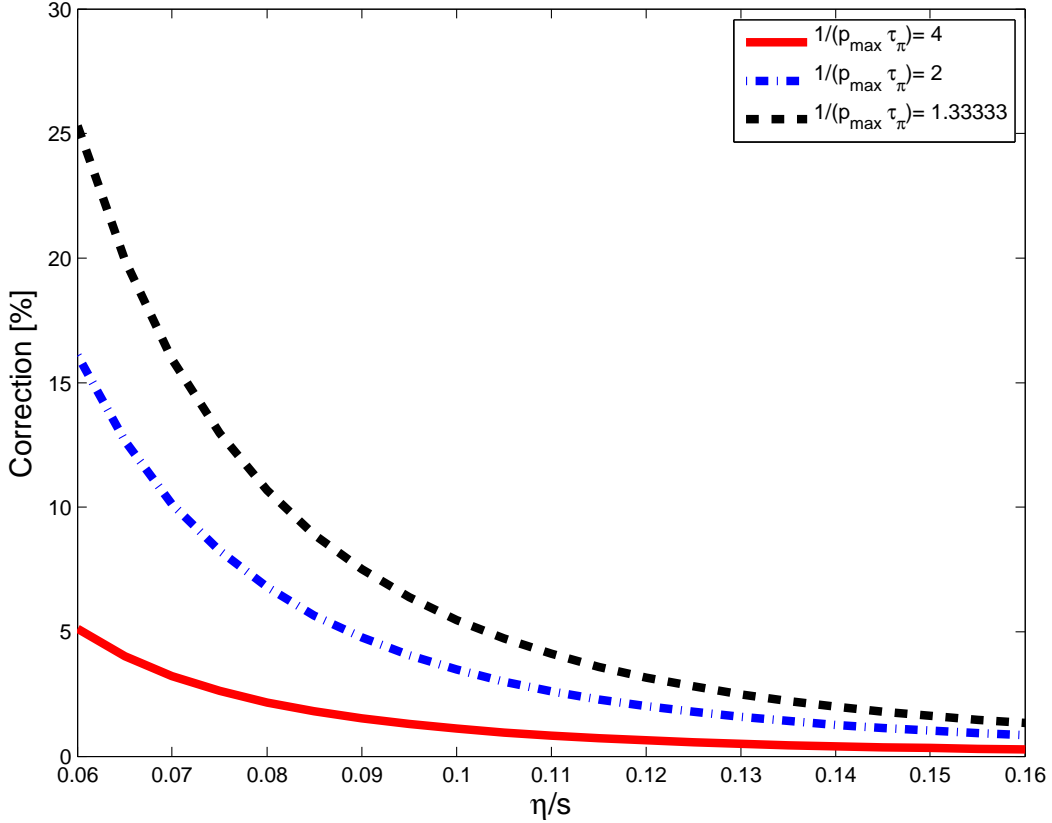}}
\caption{(Color online) Percentual correction $(\eta_c/s)\times 100/(\eta/s)$ as a function of $\eta/s$, for three different values of $p_{max}=1/(n \tau_\pi)$, with $n=4$, $n=2$ and $n=4/3$. The results correspond to $\xi = \omega_{min}/\omega_{max} = 0.88$ and $\chi = p_{max}\gamma= 1/6$ (strongly coupled theory).}
\label{f3}
\end{figure}
\end{center}

The results presented so far are based on an estimation of $p_{max}$, the maximum value of the momentum for which the gradient expansion is valid,  as $1/(2\tau_\pi)$. Since this estimation, although reasonable, is not rigurously justified, it is important and necessary to quantify the impact of different choices for $p_{max}$ on the values of $\eta_c/s$ that we obtain. To assess the strength of the conclusions that can be extracted from our results with respect to the specific value of $p_{max}$, Figure \ref{f3} shows $\eta_c/s$ as a function of $\eta/s$ for three different values of $p_{max}=1/(n \tau_\pi)$, namely $n=4$, $n=2$ and $n=4/3$, with $\xi=\omega_{min}/\omega_{max}=0.88$ (corresponding to $\eta/s=0.08$ as computed in Appendix \ref{omin}) and $\chi=p_{max}\gamma = 1/6$ (corresponding to the strongly coupled case). It is important to recall that the second order gradient expansion ceases to be valid for $p_{max}$ larger than $1/\tau_\pi$. From the results shown in Figure \ref{f3} it is seen that the correction to $\eta/s$ increases with increasing values of $p_{max}$, but the important point to note is that even for a value of $p_{max}$ quite close to $1/\tau_\pi$ (the case with $n=4/3$) the correction at $\eta/s \sim 0.08$ remains $\sim 10 \%$. Figure \ref{f3} shows that, for realistic estimates of the parameters corresponding to the strongly coupled QGP in the range $T=200$--$300$ MeV, the correction to $\eta/s$ coming from second order terms in the fluid dynamic equations is at most $\sim 10\%$.

\begin{center}
\begin{figure}
\scalebox{0.8}{\includegraphics{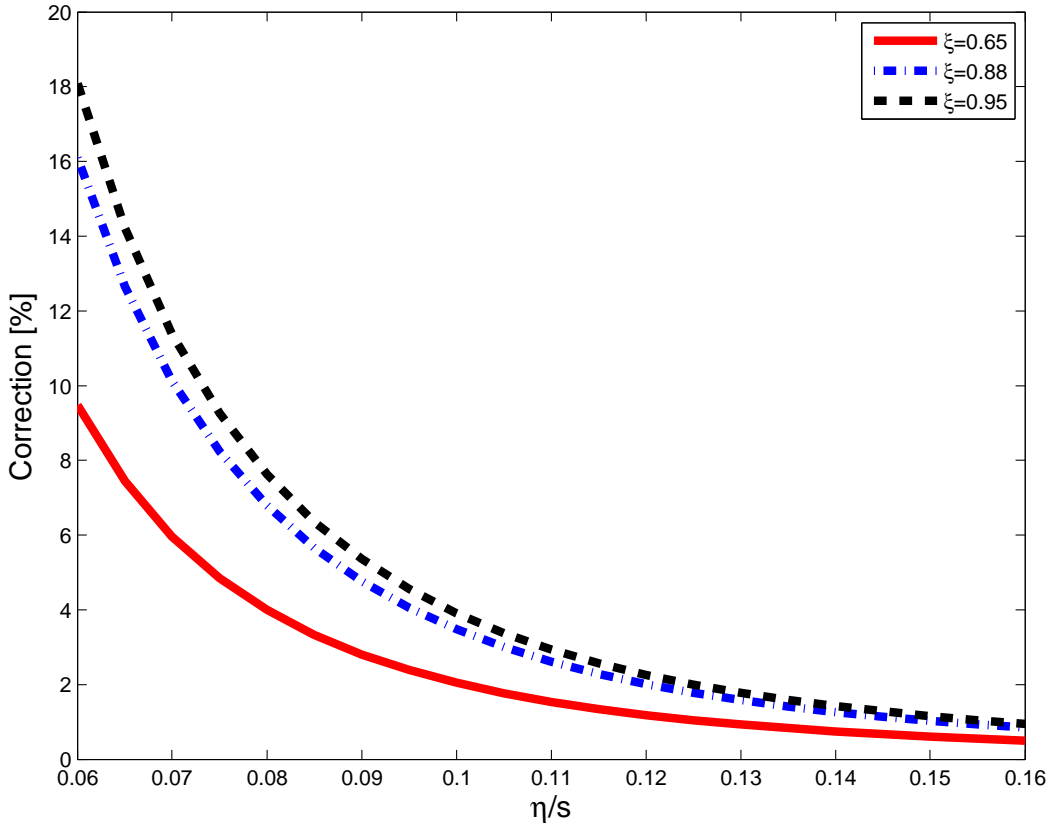}}
\caption{(Color online) Percentual correction  $(\eta_c/s)\times 100/(\eta/s)$ as a function of $\eta/s$ for $\xi = \omega_{min}/\omega_{max}=0.88$ (corresponding to the estimate for $\eta/s=0.08$ obtained in Appendix \ref{omin}), and two other close values $\xi=0.65$ and $\xi=0.95$. The value of $\chi=p_{max}\gamma$ is $1/6$, corresponding to a strongly coupled theory.}
\label{f4}
\end{figure}
\end{center} 

As discussed in Appendix \ref{omin}, here we estimate the values of $\omega_{min}$ for different values of $\eta/s$ by requiring that the equations describing the evolution of the soft modes can be linearized. In view of the fact that our approach is purely phenomenological, it is also important to quantify the effect that changing the value of $\xi$ has on the results that we obtained. To this end, Figure \ref{f4} shows $\eta_c/s$ as a function of $\eta/s$ for $\xi = \omega_{min}/\omega_{max}=0.88$, corresponding to the estimate for $\eta/s=0.08$ obtained in Appendix \ref{omin}, together with two other close values $\xi=0.65$ and $\xi=0.95$ chosen to illustrate the behavior of $\eta_c$ with $\xi$. The results correspond to the strongly coupled case with $\chi=p_{max}\gamma=1/6$. It is seen that the value of $\eta_c$ does not depend strongly on $\xi$; the variation of $\eta_c/s$ with $\xi$ is small for $\eta/s=0.08$ and negligible for $\eta/s=0.16$. These results show that the correction to $\eta/s$ for the strongly coupled theory remains bounded, giving strong support to the conclusion that $\eta_c/s < 10\%$ for the QGP under realistic conditions in the temperature range $T=200$--$300$ MeV.

Our results clearly illustrate the fact, already highlighted in \cite{kovtun}, that for a strongly coupled fluid such as the QGP formed in heavy ion collisions, thermal fluctuations actually {\it do} change the value of the effective infrared $\eta/s$, provided that the microscopic value of the ratio is small. Specifically, we have found that in strongly coupled theories with an $\eta/s$ close to the KSS value the correction due to second order terms in the fluid dynamic equations is {\it at most} $\sim 10 \%$ of the ``bare'' value, while for larger values of $\eta/s$ it is extremely small and can be safely neglected. For a weakly coupled theory the correction turns out to be much smaller than at strong coupling, and can be neglected even at $\eta/s=0.08$. These are the main results of this work. 

To put our findings into perspective, it is appropriate at this point to note that the correction $\eta_c/s$ found in Ref. \cite{kovtun} from the zeroth order term $(\rho_0 + P_0)u^\mu u^\nu$ in $T^{\mu\nu}$ is, for the same values of $p_{max}, s$, etc. as used here, roughly $50 \%$ for $\eta/s = 0.08$ (for $\eta/s=0.16$ it is negligible). The correction computed here, that comes from the second order ultrasoft term in the fluid dynamic equations, 
is at most five times smaller for $\eta/s=0.08$. 
Our results provide further support to the conclusion that, for the strongly coupled QGP created in heavy ion collisions, a value of $\eta/s=0.08$ at $T=200$--$300$ MeV seems to be ruled out by the effect of thermal fluctuations on the hydrodynamic evolution of the longest wavelength modes, but a value of $\eta/s = 0.16$ is still quite plausible. 

\section{Conclusion}
\label{summary}

In this work we compute the correction to the microscopic shear viscosity (i.e. the one computed from Kubo formula) induced from thermal fluctuations appearing in second order gradient terms in the equations of relativistic fluid dynamics. Our calculations are based on the idea of separating the hydrodynamic variables in ultrasoft and soft sectors, and determining the effect of the latter on the evolution of the former. The hard modes, which are not described by fluid dynamics, enter the conservation equations through stochastic noises. In this sense, the developments presented here fall  naturally into the relativistic fluctuating hydrodynamics framework. 

We find that the correction to $\eta/s$ is positive and inversely proportional to the square of this ratio. For the conditions prevailing in 
heavy ion collisions, we estimate the maximum correction to the viscosity--to--entropy ratio of the QGP at $T = 200$--$300$ MeV to be $\sim 10 \%$ for $\eta/s = 0.08$ and $\sim 1 \%$ for $\eta/s = 0.16$. These results indicate that for small $\eta/s$ (close to the KSS bound) the influence of soft modes on the evolution of the ultrasoft modes is significant. For larger values of $\eta/s$, the corrections are indeed very small and can be safely neglected. The correction for a weakly coupled theory is much smaller than the one found at strong coupling and can be neglected even for value of $\eta/s$ close to the KSS bound.

Our results support the conclusion found in \cite{kovtun} that thermal fluctuations have a significant impact on the value of the effective infrared $\eta/s$ corresponding to the quark--gluon plasma created in heavy ion collisions, provided that the microscopic value of $\eta/s$ is small.  Our work, and those of \cite{kap,kovtun,lub}, evidence the important fact that what could actually be extracted from data (for example by matching computed hadron observables to measurements) is an effective value of the ratio $\eta/s$ rather than the value computed microscopically. In this sense, our work gives further support to the conclusion that a value of $\eta/s = 0.08$ for the strongly coupled QGP that is created in heavy ion collisions seems impossible due to the effect of thermal fluctuations on the hydrodynamic evolution in the range $T = 200$--$300$ MeV, whereas a value of $\eta/s=0.16$ is not ruled out. 

Probably one of the most interesting things to try out is to simulate the three--dimensional evolution of the fireball created in heavy ion collisions using the equations of fluctuating hydrodynamics in order to be able to extract a value of $\eta/s$ which in principle should be closer to the real one (first steps in this direction have been taken in \cite{kap} with a boost--invariant Bjorken model). Moreover, a comparison with the values extracted from viscous hydrodynamic simulations without thermal fluctuations would allow us to directly quantify the correction to $\eta/s$ arising from the effects of such fluctuations on the hydrodynamic evolution of matter created in heavy ion collisions.

\appendix

\section{Details on the calculation of the effective shear viscosity}
\label{app}

In this appendix we describe in detail the calculation of the correction $\eta_c$ to the shear viscosity. 
We start by solving Eq. (\ref{dos}), thus obtaining 
\begin{equation}
 t^0(x) = \int dy [P]^{-1}(x-y) s_2 (y) dW(y) 
\end{equation}
so that 
\begin{equation}
 A(x) = \frac{4\rho_0 \eta}{3} \int dy [P]^{-1}(x-y)  D_0(x) S_{12}(x-y)  dW(y) 
\end{equation}
with $S_{12}(x-y)=\left\langle  s_1(x) s_2(y) \right\rangle$. Recall that $P(x)=4\rho_0 D_0/3$. 
The correction to $\eta$ comes from the term multiplying $dW$, that is to say, we would like to write $A(x)$ as $\eta_c(x) dW(x)$ for some $\eta_c(x)$. To achieve this and proceed further, we will assume that $dW(y)$ is a very slowly varying function, which allows us to approximate $A(x) \simeq \eta_c(x)dW(x)$ with 
\begin{equation}
 \eta_c (x) = \frac{4\rho_0 \eta}{3} \int dy [P]^{-1}(x-y)  D_0(x) S_{12}(x-y) 
\label{etacx}
\end{equation}
The approximation that $dW$ be a slowly-varying function is appropriate for our aim of providing an estimative upper bound on $\eta_c$, and in any case it can be regarded as a definition of $\eta_c$ that is valid for all practical purposes. 

Going over to momentum space, we can write
\begin{equation}
 \eta_c (k^\mu) = -\frac{\eta}{(2\pi)^4} \int d^4\lambda ~\frac{\lambda^0}{k^0-\lambda^0} S_{12}(k^\mu-\lambda^\mu)
\label{etaaux}
\end{equation}
or, putting $\sigma^\mu = k^\mu - \lambda^\mu$, 
\begin{equation}
 \eta_c (k^\mu) = \frac{\eta}{(2\pi)^4} \int d^4\sigma~ \frac{k^0-\sigma^0}{\sigma^0} S_{12}(\sigma^\mu)
\label{etack}
\end{equation}
It is important to remark that the frequency integrations in Eqs. (\ref{etaaux}) and (\ref{etack}) run over all (positive and negative) values of the frequency. Since $S_{12}$ is even in $\sigma^0$ (it must due to the invariance of the equilibrium state), we can change the integration limits in Eq. (\ref{etack}) to $\sigma^0 \in (0,\infty)$. An immediate consequence of this is that the terms $\propto k^0$ drop out when integrated, and hence need not be considered further in what follows. 

Assuming that the noises satisfy the fluctuation--dissipation theorem, the correlator of velocity fluctuations can be readily computed \cite{landau,esteban-fluc,librofluc,fox,libro,kovtun}. It is given by
\begin{equation}
 S_{12}(\sigma^\mu) = -\frac{2T\gamma}{(\rho_0 + P_0)}\frac{\sigma^2}{(\sigma^0)^2 + \gamma^2 \sigma^4} + \frac{8T\gamma}{3(\rho_0+P_0)}\frac{(\sigma^0)^2 \sigma^2}{[(\sigma^0)^2-\sigma^2/3]^2+(4\gamma \sigma^2 \sigma^0/3)^2}
\label{s12}
\end{equation}
with $\sigma = \sigma^i \sigma_i$ and $\gamma=\eta/(\rho_0 + P_0) = \eta/sT$ ($s$ is the entropy density). The first and second terms correspond to shear and sound waves, respectively. 

We will compute the contributions to $\eta_c$ coming from shear and sounds waves separately. Let us start with the shear waves. 
We get  
\begin{equation}
 \eta_c = \frac{T\gamma^2}{\pi^3} \int_{\omega_{min}}^{\omega_{max}} d\sigma^0~ \frac{1}{(\sigma^0)^2} 
\int_0^{p_{max}}  \frac{\sigma^4}{1+b \sigma^4} d\sigma
\label{etac1}
\end{equation}
where we have defined $b=(\gamma/\sigma^0)^2$. The rationale for introducing in Eq. (\ref{etac1}) the upper and lower bounds on the frequency $\sigma^0$ is discussed below.

As mentioned in Section \ref{fluc}, we assume that only soft and ultrasoft momenta contribute to $\eta_c$, since these are the modes actually described by fluid dynamics. This means that the integral must be done over modes satisfying $(\sigma^0,\sigma^i) \leq (\omega_{max},p_{max})$, where $(\omega_{max},p_{max})$ is the value of the four momentum separating soft and hard modes as discussed in Section \ref{fluc}.
The reason for introducing $\omega_{min} \neq 0$ in Eq. (\ref{etac1}) is to ensure that the equations for the soft modes can be linearized, as discussed in detail in Appendix \ref{omin}. Estimative values of both cutoffs $\omega_{max}$ and $\omega_{min}$ are discussed in Section \ref{num}. 

We now return to Eq. (\ref{etac1}). Assuming momentum isotropy the spatial integral becomes
\begin{equation}
\int_0^{p_{max}}  \frac{\sigma^4}{1+b \sigma^4} d\sigma = \frac{p_{max}(\sigma^{0})^2}{\gamma^2} 
- \frac{1}{4\sqrt{2}}\bigg(\frac{\sigma^0}{\gamma}\bigg)^{5/2}\bigg[ \ln \bigg(\frac{A_{1}}{A_{2}}\bigg) + 2\textrm{arctan} \bigg(\frac{\sqrt{2}\alpha p_{max}}{\alpha^2 - p^2_{max}}\bigg)  \bigg]
\end{equation}
with 
\begin{equation}
A_{1,2}=p^2_{max}\pm \sqrt{2}\alpha p_{max} + \alpha^2
\end{equation}
and 
\begin{equation}
\alpha=\sqrt[4]{1/b}=\sqrt{\sigma^0/\gamma} 
\end{equation}

We shall obtain the general expression for $\eta_c$ for $k^0 \neq 0$. Defining 
\begin{equation}
\eta_c^{(1)} = \frac{T\gamma^2}{\pi^3}  \int_{\omega_{min}}^{\omega_{max}} d\sigma^0~  \frac{p_{max}}{\gamma^2} 
\end{equation}
\begin{equation}
\eta_c^{(2)} = -\frac{T}{2^{5/2}\pi^3\gamma^{1/2}} \int_{\omega_{min}}^{\omega_{max}} d\sigma^0~\sqrt{\sigma^0} \ln \bigg( \frac{A_1}{A_2}\bigg)
\end{equation}
and 
\begin{equation}
\eta_c^{(3)} =  -\frac{T}{2^{3/2}\pi^3\gamma^{1/2}} \int_{\omega_{min}}^{\omega_{max}} d\sigma^0~ \sqrt{\sigma^0} \textrm{arctan} \bigg( \frac{\sqrt{2} p_{max}\alpha}{\alpha^2-p^2_{max}}\bigg)
\end{equation}
the first contribution becomes 
\begin{equation}
\eta_c^{(1)} = \frac{T p_{max}}{\pi^3} [\omega_{max}-\omega_{min}]
\label{etac11}
\end{equation}

In terms of $\Phi=\alpha/p_{max}$ we have
\begin{equation}
\eta^{(2)}_c = -\frac{T p_{max}}{2^{3/2}\pi^3}\int_{\Phi_{min}}^{\Phi_{max}} d\Phi~\gamma p^2_{max}\Phi^2 
\ln \bigg(\frac{1+\sqrt{2}\Phi + \Phi^2}{1-\sqrt{2}\Phi + \Phi^2} \bigg)
\end{equation}
and
\begin{equation}
\eta^{(3)}_c = -\frac{T p_{max}}{2^{5/2\pi^3}}\int_{\Phi_{min}}^{\Phi_{max}} d\Phi~ \gamma p^2_{max}\Phi^2 \textrm{arctan} \bigg(\frac{\sqrt{2}\Phi}{\Phi^2-1} \bigg)
\end{equation}

Performing the integration we get 
\begin{equation}
\begin{split}
\eta^{(2)}_c &= - \frac{T\gamma p^3_{max}}{3\sqrt{2}} \bigg[ -\sqrt{2}\ln\bigg(\frac{1+\Phi_{max}^4}{1+\Phi_{min}^4} \bigg)  
+2\Phi_{max}^3\ln\bigg(\frac{1+\sqrt{2}\Phi_{max}+\Phi_{max}^2}{1-\sqrt{2}\Phi_{max}+\Phi_{max}^2} \bigg) \\
&- 2\Phi_{min}^3\ln\bigg(\frac{1+\sqrt{2}\Phi_{min}+\Phi_{min}^2}{1-\sqrt{2}\Phi_{min}+\Phi_{min}^2} \bigg) 
+2\sqrt{2} \textrm{arctan}\bigg( \frac{\Phi^2_{max}+\Phi^2_{min}}{\Phi^2_{max}\Phi^2_{min}-1}\bigg)  \\
&+ 2\sqrt{2}(\Phi^2_{max}-\Phi^2_{min}) \bigg] 
\end{split}
\label{etac2}
\end{equation}
and 
\begin{equation}
\begin{split}
\eta^{(3)}_c &= -  \frac{T\gamma p^3_{max}}{3\sqrt{2}} \bigg[ \sqrt{2}\ln\bigg(\frac{1+\Phi_{max}^4}{1+\Phi_{min}^4} \bigg)  
+2\sqrt{2} \textrm{arctan}\bigg( \frac{\Phi^2_{max}+\Phi^2_{min}}{\Phi^2_{max}\Phi^2_{min}-1}\bigg)  \\
&+ 2\sqrt{2}(\Phi^2_{max}-\Phi^2_{min}) +4\Phi_{max}^3 \textrm{arctan}\bigg( \frac{\sqrt{2}\Phi^2_{max}}{\Phi^2_{max}-1}\bigg) 
-4\Phi_{min}^3 \textrm{arctan}\bigg( \frac{\sqrt{2}\Phi^2_{min}}{\Phi^2_{min}-1}\bigg)   \bigg] 
\end{split}
\label{etac3}
\end{equation}

We now go over to compute the contribution to $\eta_c$ coming from the sound waves. We have
\begin{equation}
\eta_c = - \frac{4T\gamma^2}{3\pi^3} \int_{\omega_{min}}^{\omega_{max}} d\sigma^0~ (\sigma^0)^2 \int_0^{p_{max}} d\sigma~ 
\frac{\sigma^4}{[(\sigma^0)^2-\sigma^2/3]^2+(4\gamma \sigma^2 \sigma^0/3)^2}
\label{etasound}
\end{equation}
To provide an analytic result we shall be concerned only with the case of small $\gamma$. The idea is that the behavior of Eq. (\ref{etasound}) for $\gamma\to 0$ is dominated by the singularity at $(\sigma^0)^2=\sigma^2/3$. To isolate the singular behavior, we write 
\begin{equation}
\eta_c = - \frac{4T\gamma^2}{3\pi^3}  \int_{\omega_{min}}^{\omega_{max}} d\sigma^0~ \int_0^{p_{max}} d\sigma~ 
\frac{(\sigma^0)^2 \sigma^4}{[(\sigma^0)^2-\sigma^2/3+4i\gamma \sigma^2 \sigma^0/3][(\sigma^0)^2-\sigma^2/3-4i\gamma \sigma^2 \sigma^0/3]}
\end{equation}
Completing the squares, the integrand reads
\begin{equation}
\frac{(\sigma^0)^2 \sigma^4}{[(\sigma^0+2i\gamma \sigma^2/3)^2-\sigma^2/3+4\gamma^2 \sigma^4/9][(\sigma^0-2i\gamma \sigma^2/3)^2-\sigma^2/3+4\gamma^2 \sigma^4/9]}
\end{equation}
Since we are interested in the small $\gamma$ behavior, we neglect the $\gamma^2$ terms so 
\begin{equation}
\eta_c = - \frac{4T\gamma^2}{3\pi^3} \int_{\omega_{min}}^{\omega_{max}} d\sigma^0~  \int_0^{p_{max}} d\sigma~ 
\frac{(\sigma^0)^2 \sigma^4}{[(\sigma^0+2i\gamma \sigma^2/3)^2-\sigma^2/3][(\sigma^0-2i\gamma \sigma^2/3)^2-\sigma^2/3]}
\end{equation}
We assume $\omega_{min}$ is small enough and $\omega_{max}$ is large enough that we can compute the integral by closing the contour. We choose to close it from above. We thereby pick up the poles at $2i\gamma \sigma^2/3\pm\sigma/\sqrt{3}$. We get 
\begin{equation}
\begin{split}
\eta_c = - \frac{4iT\gamma^2}{3\pi^2} \int_0^{p_{max}} d\sigma~ 
&\bigg[\frac{(2i\gamma \sigma^2/3+\sigma/\sqrt{3})^2 \sigma^4}{[(\sigma/\sqrt{3}+4i\gamma \sigma^2/3)^2-\sigma^2/3][2\sigma^2/3]}\\
&+ \frac{(2i\gamma \sigma^2/3-\sigma/\sqrt{3})^2 \sigma^4}{[(-\sigma/\sqrt{3}+4i\gamma \sigma^2/3)^2-\sigma^2/3][-2\sigma^2/3]} \bigg]
\end{split}
\end{equation}
We keep only the leading $\gamma$ terms, whereby the integral becomes trivial and gives $\approx p_{max}^3/\gamma$. The final result is 
\begin{equation}
 \eta_c = -\frac{T\gamma p_{max}^3}{6\pi}
\label{etac-sound}
\end{equation}

\section{Estimating $\omega_{min}$}
\label{omin}

In this appendix we describe the procedure used to estimate the value of $\omega_{min}$, the frequency that divides ultrasoft from soft modes. The idea is to find $\omega_{min}$ by requiring that the soft mode equations can be linearized \cite{libro,blaizot}. 

The condition for a mode with wavevector $k$ to be linearizable is
\begin{equation}
k\left\langle v^2 \right\rangle \leq \nu k^2 \sqrt{\left\langle v^2 \right\rangle} 
\end{equation}
where
\begin{equation}
\nu=\frac{\eta}{\rho_0} = \frac{4\eta}{3sT}
\end{equation}
and $\left\langle v^2 \right\rangle$ is the thermal average of the square of the soft mode velocity. This thermal average can be estimated from the magnitude of the shear correlator $S_{12}$ given in Eq. (\ref{s12}). For our purpose of estimating $\omega_{min}$ it is enough to consider only the shear part of the correlator. We get
\begin{equation}
\left\langle v^2 \right\rangle = \frac{8\pi \eta}{asT^4} \int_k^{p_{max}} d\sigma^0~ \int_k^{p_{max}}d\sigma~ \frac{\sigma^4}{(\sigma^0)^2+(\eta/s)^2(\sigma^4/T^2)}
\end{equation}
where we have used that $s=aT^3$ and set $\omega_{max}=p_{max}$. For the $\sigma^0$ integral, we call
\begin{equation}
\sigma^0 = \frac{\eta}{s}\frac{\sigma^2}{T} \textrm{tan} \phi
\end{equation}
so we get
\begin{equation}
\left\langle v^2 \right\rangle = \frac{8\pi}{aT^3} \int_k^{p_{max}} d\sigma~ \sigma^2 \bigg[\textrm{arctan}\bigg(\frac{p_{max}T}{(\eta/s)\sigma^2} \bigg) - \textrm{arctan}\bigg(\frac{kT}{(\eta/s)\sigma^2} \bigg)\bigg]
\end{equation}

If $\eta/s$ is small, the arguments of both arctans are large, and we may approximate 
\begin{equation}
\textrm{arctan}\bigg(\frac{1}{\delta} \bigg) \approx \frac{\pi}{2} - \delta
\end{equation}
to get 
\begin{equation}
\left\langle v^2 \right\rangle = \frac{8\pi}{5aT^4}\frac{\eta}{s} \bigg(\frac{1}{k}-\frac{1}{p_{max}} \bigg) (p^5_{max}-k^5) \approx \frac{8\pi}{aT^4}\frac{\eta}{s}p^2_{max} (p_{max}-k)^2
\end{equation}

Putting all together we get the condition for soft modes as
\begin{equation}
\bigg(1-\frac{k}{p_{max}}\bigg) \leq y \bigg( \frac{\eta}{s}\bigg)^{1/2} \frac{T k}{p^2_{max}}
\end{equation}
with $y$ some constant of order one. The boundary lies at
\begin{equation}
\xi = \frac{\omega_{min}}{p_{max}} = \frac{1}{1+y(\frac{\eta}{s})^{1/2} \frac{T}{p_{max}}}
\label{xiapp1}
\end{equation}
This is our final result. For simplicity, in order to estimate the value of $\xi$ we will set $y=1$. For later convenience, we will rewrite Eq. (\ref{xiapp1}) in terms of $\chi=p_{max}\eta/(sT)$. We have
\begin{equation}
\xi = \frac{1}{1+(\frac{\eta}{s})^{3/2}\frac{1}{\chi}}
\label{xiapp2}
\end{equation}
For a strongly coupled theory with $\chi=1/6$ we get from Eq. (\ref{xiapp2}) that $\xi =0.88$ for $\eta/s=0.08$ and $\xi=0.72$ for $\eta/s=0.16$. For a weakly coupled theory with $\chi=1/12$  we get $\xi =0.78$ for $\eta/s=0.08$ and $\xi=0.56$ for $\eta/s=0.16$.

For completeness, we briefly discuss the approach followed in \cite{kovtun} to estimate the value of $\omega_{min}$, which is different from the one adopted here. There it was found that the contribution of the quadratic term $(\rho_0 + P_0)u^\mu u^\nu$ to the retarded correlation function of two stress energy tensor operators $T^{xy}$ gives rise to a nonanalytic term that goes like $\omega^{3/2}$. The value of $\omega_{min}$ can be found by comparing the standard term $\eta \tau_\pi \omega^2$ to the nonanalytic correction. For $\omega < \omega_{min}$, second order fluid dynamics ceases to be valid because the gradient expansion breaks down, because the nonanalytic correction becomes larger than the standard term. In our notation, the result can be expressed as follows (see Eq. 4.8 of \cite{kovtun})
\begin{equation}
\xi \simeq \frac{4 \chi^2 10^{-4}}{a^2} \bigg( \frac{\eta}{s}\bigg)^{-6}
\label{xiprime}
\end{equation}
From Eq. (\ref{xiprime}) we get, for the QGP with $a=10$ and $\chi=p_{max}\gamma=1/6$, that $\xi=0.42$ for $\eta/s = 0.08$ and $\xi = 0.007$ for $\eta/s=0.16$. These values are smaller than the ones we obtain from requiring that the soft mode equations be linearizable. 

\begin{acknowledgments}
This work has been supported in part by ANPCyT, CONICET and UBA under project UBACYT X032 (Argentina), and by FAPESP (Brazil).
\end{acknowledgments}

\end{document}